\documentclass[12pt,preprint]{aastex}

\newcommand{\dd}{\mathrm{d}}
\newcommand{\bp}{\bar{p}}
\newcommand{\bw}{\bar{w}}
\newcommand{\pl}{P_{\mathrm{l}}}
\newcommand{\rj}{\;R_{\mathrm{jup}}}

\shorttitle{The Polarization Signature of ESPs Transiting Cool Dwarfs}
\shortauthors{Carciofi \& Magalh\~aes}

\begin{document}
\title{The Polarization Signature of Extrasolar Planets Transiting Cool Dwarfs}


   \author{A.C. Carciofi
          and
          A.M. Magalh\~aes
          }

     \affil{Instituto de Astronomia, Geof\'{i}sica e Ci\^encias Atmosf\'ericas, Rua do Mat\~ao 1226, Cidade Universit\'aria, S\~ao Paulo, SP, BRAZIL, 05508-900}

    \email{carciofi@usp.br, mario@astro.iag.usp.br}

\begin{abstract}
 We investigate the linear polarization in the light of extrasolar planetary systems that may arise as a result of an occultation of the star by a transiting planet. Such an occultation breaks any spherical symmetry over the projected stellar disk and thus results in a non-vanishing linear polarization. This polarization will furthermore vary as the occultation progresses. We present both analytical and numerical results for the occultation of G-K-M-T dwarf stars by planets with sizes ranging from the one of Earth to two times the size of Jupiter.  We find that the occultation polarization may result in an observable signal and provide additional means to characterize various parameters of the system. 
A particularly interesting result is that, for the later spectral types (i.e., smaller stellar radii), this polarization signature may be observable even for Earth-like planets. This suggests polarization as a possible tool to detect such planets.  Departs from symmetry around mid-transit in the time dependence of the polarization signature may provide an estimate of the orbital eccentricity.
\end{abstract}


\section{Introduction}
In the wake of the first discovery of an extrasolar planet (ESP) orbiting a solar-type star by \citet{may95}, over 150 ESPs have been found to date \citep[see][for a review]{fer05}. The vast majority of the discoveries have used the radial velocity method, in which high accuracy and precision Doppler measurements of the stellar spectrum yield the influence of the planet on the stellar motion. Among these ESPs, one (HD209458b) has been observed to transit its star. As of May 2005 another 6 ESPs have been discovered photometrically through transits \footnote{http://www.obspm.fr/encycl/encycl.html}.  Recently, \citet{cha05} and \citet{dem05} reported the first direct measurement of the radiation of an ESP using the Spitzer Space Telescope.

In addition to estimates of the ESP's period and, if the stellar mass is known, the semi-major axis of the planet orbit, the radial velocity curve provides a value for the minimum mass of the ESP, $m \sin i$, and the eccentricity of the orbit. For a transiting planet, the inclination $i$ can be estimated, and hence the ESP's mass, in addition to the planet's size and atmospheric contents \citep[e.g.,][]{alo04}.

This lopsided balance in favor of the radial velocity method as a means to detect ESPs may be turned around in the near future with the forthcoming satellite missions. As an example, the COROT mission is expected to detect hundreds of new planets using the photometry method to measure occultations by transiting ESPs with micromagnitude accuracy. Such accuracy should also allow the detection of non--transiting ESPs through the added variation of their reflected light along their orbit \citep{hou03}.

The use of polarimetry as a tool to detect and study ESPs has been already expounded 
\citep{sea00,saa03,hou03,sta04}.
The main motivation in those works was precisely that the ESP will contribute to the light beam with reflected starlight, at least in the optical and near infrared up to $3\;\mu \mathrm{m}$  \citep{sta04}. In the case of an \emph{unresolved} star-ESP system, the observed polarization is expected to be rather small, $\approx 10^{-5} (=10^{-3}$\%) or less. This is because the unpolarized contribution from the star will be so much greater than the planet's reflected light \citep{sea00, saa03}, the degree of polarization being essentially given by the ratio of reflected to total (= polarized + unpolarized) light. The direct detection of a \emph{resolved} ESP via its reflected light has been detailed by \citet{sta04}. Of course, for a planetary atmosphere the term \emph{scattering} (by molecules and dust particles, for instance) of starlight might be in general more appropriate than {\it reflection}, at least for Jupiter-like or gaseous ESPs.

Nevertheless, polarimetry is a relative measurement and high accuracy is possible. Historically, the first mention of using highly accurate polarimetry for ESP detection seems to have been made by Kemp (oral communication) who built an equipment capable of $\approx 10^{-7}$ accuracy, given enough photons \citep{kem87}. \citet{hou03} report on a polarimeter being built with a sensitivity better than $10^{-6}$ and aimed at detecting unresolved ESPs. At the same time, an equipment combining polarimetry, spectroscopy and adaptive optics 
\citep[CHEOPS,][]{fel03}
has been proposed, which should achieve $10^{-5}$ fractional accuracy. All these are or will be ground-based equipment; a comparable photometric accuracy can be typically obtained \emph{only} from space.
Light emitted by a spherical star will present a very small polarization. \citet{kem87} have measured a $V$-band linear polarization of less than  $2 \times 10^{-7}$ for the integrated solar disk, for example. However, when we resolve the solar disk, a polarization variation from center to limb will ensue  \citep{ler00}. The center-to-limb variation in the polarization across any stellar disk originates from an existing contribution of a scattering opacity to the total opacity in the atmosphere \citep{har69}. In hot stars, electron (Thompson) scattering is one such opacity. In the case of cool stars, Rayleigh scatter     ing by atomic and molecular hydrogen and atomic helium provide the main source of scattering opacity. 

Barring any other source of asymmetry over the stellar disk (e.g., spots, flattening by rotation, etc.), the polarization pattern of a spherical star at a given limb angle will be tangential to the radial direction and the polarization value itself will typically increase from center to limb. This is because we tend to see more scattered, hence polarized, light towards the stellar limb. When a planet transits the star, it will then break this centro-symmetric pattern and cause a net linear polarization if we integrate the polarization across the stellar disk.

In this paper we model and estimate the {\it occultation polarization} (henceforth OP) that results from the transit of an ESP. We consider stellar spectral types from G through T (brown dwarfs) and planet radii ranging from one Earth radius to two Jupiter radii. We first describe the problem and consider analytical approximations for the OP in Sect.~\ref{basic}. In Sect.~\ref{results} we describe the simulation results for cool dwarfs and discuss the observational prospects as well as what can be learned from such observations. Finally, in Sect.~\ref{conclusions} we summarize our results.

\section{Basic Equations \label{basic}}

\begin{figure}
  \plotone{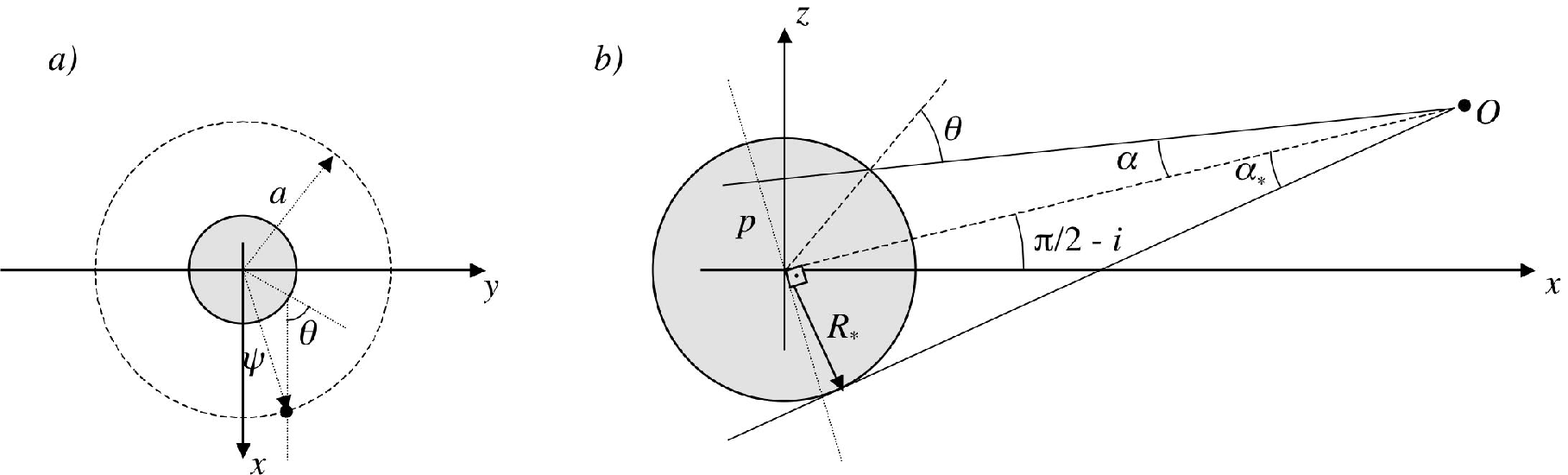}
   \caption{Transity geometry. The observer's direction is in the $xz$ plane and has an angle $i$ with respect to the $z$-axis.}
              \label{Geometria}%
    \end{figure}

Consider the geometry shown in Fig.~\ref{Geometria}-a, representing a planet orbiting a star  with a circular orbit of radius $a$ in the $xy$ plane. At a given time, the position of the planet is given by its azimuth $\psi$, measured from the $x$-axis. 
The direction to the observer is in the $xz$ plane and has an inclination angle $i$ with respect to the $z$-axis. The observer is at a distance $d$ from the star (Fig.~\ref{Geometria}-b).

Using this geometry we derive the expressions for the observed stellar flux and Stokes parameters $Q$ and $U$.

\subsection{Flux}

The stellar flux at $O$ at a given wavelength is given by
\begin{equation}
F_\star = \int_{0}^{2\pi}\dd\phi
\int_0^{\alpha_\star} I_0 f(\mu)\cos\alpha\sin\alpha\,\dd\alpha, 
\label{fluxdefinition}
\end{equation}
where $I_0$ is a constant, $\alpha_\star \equiv \arcsin \left( R_\star/d \right)$ is the angle at which a ray from $O$ is tangent to the star, $\mu\equiv\cos\theta$ is the angle between the normal to the surface and the direction to the observer  (Fig.~\ref{Geometria}-b) and  $f(\mu)$ is the limb darkening law,
defined so that
\begin{eqnarray*}
\int_0^1  \mu  f(\mu) d\mu \equiv \frac{1}{2}.
\end{eqnarray*}
As will be evident later, a more convenient integration variable in Eq.~(\ref{fluxdefinition}) is $\bp$, the normalized impact parameter of a ray from $O$, given by
\begin{eqnarray*}
\bp\equiv\frac{p}{R_\star} = \frac{d}{ R_\star} \frac{\sin\alpha}{\cos\alpha},
\end{eqnarray*}
In the limit $\alpha \rightarrow 0$ (or $d >> R_\star$), $\bp$ becomes
\begin{eqnarray*}
\bp = \frac{d}{ R_\star} \sin\alpha,
\end{eqnarray*}
and equation~(\ref{fluxdefinition}) takes the form
\begin{equation}
F_\star = I_0\frac{R_\star^2}{d^2} \int_{0}^{2\pi}\dd\phi
\int_0^1 \bp f(\mu)\dd \bp. 
\label{fluxintermsofp}
\end{equation}
From the relation $\bp=\sqrt{1-\mu^2}$, also valid in the limit $\alpha \rightarrow 0$, we have
$F_\star = \pi I_0 \left({R_\star}/{d}\right)^2,$
which is the familiar inverse square law for the stellar flux.

Let us now calculate the flux absorbed by a planet with phase $\psi$, observed from an inclination $i$. The projected coordinates of the planet center in the plane of the sky are, in the limit $\alpha \rightarrow 0$,
\begin{eqnarray}
\bp_0 = \frac{a}{ R_\star} \sqrt{\sin^2\psi + \cos^2\psi\cos^2i}, \label{p_psi_i}\\
\phi_0 = \arctan \left( \frac{\cos i}{\tan\psi}\right ) \label{phi_psi_i}.
\end{eqnarray}

The flux absorbed by the planet is given by the integral of $I_0 f(\mu)$ over the circle projected by the planet on the stellar disk. From Eq.~(\ref{fluxintermsofp}) we have
\begin{eqnarray*}
F_\mathrm{abs} (\bp_0,\phi_0)= I_0\frac{R_\star^2}{d^2}
\int_{0}^{2\pi}\dd\phi^\prime
\int_0^{\bw} \bp^\prime f(\bp^\prime,\phi^\prime)\dd \bp^\prime,
\label{AbsorbedFlux}
\end{eqnarray*}
where $\bw \equiv R_\mathrm{p}/R_\star$ and $\bp^\prime$ and $\phi^\prime$ are polar coordinates of a system whose origin is $(\bp_0,\phi_0)$.

A more useful quantity, which does not depend on $d$ or $I_0$, is the observed normalized flux, defined as
\begin{eqnarray}
F(\bp_0,\phi_0) \equiv 1 - \frac{F_\mathrm{abs}(\bp_0,\phi_0)}{F_\star} = 1-\frac{1}{\pi} 
\int_{0}^{2\pi}\dd\phi^\prime
\int_0^{\bw} \bp^\prime f(\bp^\prime,\phi^\prime)\dd \bp^\prime. \label{fobs}
\end{eqnarray}

\subsection{Polarization}

We assume the limb polarization is a function only of $\mu$, $P = P\left(\mu\right)$,  
and the polarization direction is centrosymmetric with respect to the direction between the center of the star and the observer. Thus, for a given $\bp$ and $\phi$, the Stokes parameters are given by
\begin{eqnarray}
Q(\bp,\phi) = I_0 f(\mu)P(\mu)\cos(2\phi), \label{qdef}\\
U(\bp,\phi) = -I_0 f(\mu)P(\mu)\sin(2\phi).\label{udef}
\end{eqnarray}
It follows from Eq.~(\ref{AbsorbedFlux}), (\ref{qdef}) and (\ref{udef}) that the observed normalized Stokes parameters, $q=Q/F_\mathrm{abs}$ and $u=U/F_\mathrm{abs}$, are
\begin{eqnarray}
q(\bp_0,\phi_0) = \frac{1}{\pi} 
\int_{0}^{2\pi}\dd\phi^\prime
\int_0^{\bw} \bp^\prime f(\bp^\prime,\phi^\prime)
P(\bp^\prime,\phi^\prime)\cos(2\phi)
\dd \bp^\prime, \label{qobs}\\
u(\bp_0,\phi_0) = -\frac{1}{\pi} 
\int_{0}^{2\pi}\dd\phi^\prime
\int_0^{\bw} \bp^\prime f(\bp^\prime,\phi^\prime)
P(\bp^\prime,\phi^\prime)\sin(2\phi)
\dd \bp^\prime, \label{uobs}
\end{eqnarray}
Note that in the above expressions, the argument of the sine and cosine is $2\phi$, without the prime. Note, also, that the expressions above for the observed flux and Stokes parameters are not valid for the ingress and egress phases of the transit.

\subsection{Analytical approximations \label{ana}}

In this section we derive a simple analytical approximation for $q$ and $u$, valid when $\bw \ll 1$. In this case, the argument of the integrals in Eqs.~(\ref{qobs}) and (\ref{uobs}) is approximately constant
\begin{eqnarray*}
f(\bp^\prime,\phi^\prime) \approx f(\bp_0), \\
P(\bp^\prime,\phi^\prime) \approx P(\bp_0),
\end{eqnarray*}
and the observed normalized Stokes parameters take the following simple form
\begin{eqnarray*}
q(\bp_0,\phi_0) \approx \bw^2 f(\bp_0) P(\bp_0) \cos(2\phi_0), \\
u(\bp_0,\phi_0) \approx -\bw^2 f(\bp_0) P(\bp_0) \sin(2\phi_0). 
\end{eqnarray*}
The observed OP in this approximation is given by
\begin{equation}
p(\bp_0) = \left[ q(\bp_0)^2 + u(\bp_0)^2 \right]^{1/2} \approx \bw^2 f(\bp_0) P(\bp_0). 
\end{equation}The polarization depends only on the impact parameter, because of the centro-symmetric pattern of the stellar limb polarization. 

Let us now obtain an expression for the maximum OP during the transit. Because the limb polarization usually grows towards the limb, we expect that the maximum polarization occurs near the interior contact of either the ingress or egress phases. The impact parameter of the planet center at this phase is $\bp_0 = 1-\bw$, hence the maximum polarization is
\begin{equation}
p_{\mathrm{max}} \approx \bw^2 f(1-\bw) P(1-\bw). \label{pmax}
\end{equation}

The above expression shows that the maximum polarization is controlled by two quantities with opposite effects. The first quantity is the ratio of the solid angles subtended by the planet and the star ($\bw^2$); the larger this ratio, the larger the flux absorbed by the planet and the larger the polarization.
The second quantity is the limb polarization at the impact parameter of the planet center at the contact. Here, the larger the size of the planet, the lower the impact parameter and, thus, the lower the polarization.
As we discuss in the following section, the existence of these two competing effects has important consequences on the maximum OP produced.

\section{Numerical Results \label{results}}

\subsection{Monte Carlo Code}

In the following sections, we present numerical polarization results for representative cases of G, K, M and T dwarfs. 
We used a Monte Carlo integrator for solving Eqs.~(\ref{qobs}) and (\ref{uobs}).
For a given configuration of the star-planet system (i.e., $R_\star$, $R_\mathrm{p}$, $a$,  and $\psi$), the procedure adopted is as follows.
A photon packet (PP) is emitted in a random position on the stellar surface and in the direction towards the observer. The PPs are given the following weight
\begin{equation}
\epsilon_i = \frac{f(\mu)L_\star}{N},
\end{equation}
where  $L_\star$ is the stellar luminosity. The above expression means simply that the weight is proportional to the probability of the PP being emitted towards the observer. In addition to the weight, the PP is characterized by its linear polarization
\begin{eqnarray}
Q_i = \epsilon_i P\left(\mu\right) \cos(2\phi), \\
U_i = -\epsilon_i P\left(\mu\right) \sin(2\phi).
\end{eqnarray}

If the PP is emitted in the hemisphere opposite to the one facing the observer it is given weight zero, because it can not reach the observer. If, on the contrary, the PP is emitted in the hemisphere facing the observer, it is checked whether its trajectory crosses the planet. If it does, the PP will be absorbed by planet and, thus, is also given weight zero.

This procedure is repeated $N$ times and, at the end of the simulation, the flux and Stokes parameters will be given by the following estimators
\begin{eqnarray}
F_\psi = \sum_N w_i, \\
q_\psi = \frac{\sum_N Q_i}{F_\psi}, \\
u_\psi = \frac{\sum_N U_i}{F_\psi},
\end{eqnarray}
where the subscript $\psi$ indicates the dependence of the estimators with planet position. $F$, $q$ and $u$ are calculated for several planet positions in order to simulate the temporal progression of the transit. 

The above procedure is extremely simple to implement and it has the advantage of avoiding dealing with the complications that arise on the ingress and egress phases of the transit, for which only part of the planet disk blocks stellar light. In addition, other physical effects (such as stellar spots) can be easily implemented.

\subsection{Limb Darkening and Polarization}

For the limb darkening law we use the formula proposed by \citet{cla00}
\begin{equation}
f(\mu) = 1 - a_1(1-\mu^{1/2}) - a_2(1-\mu) - a_3(1-\mu^{3/2}) - a_4(1-\mu^2),
\end{equation}
where the $a_i$ are constants that depend on wavelength, spectral type, surface gravity and metallicity.
In Table~\ref{tab1} we list the $a_i$ we used for each spectral class, as well as the radius and effective temperatures adopted. Solar metallicities are assumed.

In contrast to the limb darkening, which is relatively well-known for normal stars, the limb polarization is well-known only for the Sun, for which it was extensively studied both observationally and theoretically \cite[e.g.,][]{flu99,ler00, fau01, fau03}. For other spectral types, however, to our knowledge only a handful of theoretical studies has been made  \citep{har69,har70,mag86,col88}. 

For the Sun, we adopt a limb polarization of the following form \citep{flu99}
\begin{equation}
P(\mu) = \pl \frac{1-\mu^2}{1+k\mu},
\label{limbp}
\end{equation}
where $k$ is a measure of the width of the polarization profile (i.e., the larger the $k$, the fastest the polarization drops from the value at the limb). Both $\pl$ and $k$ are a function of wavelength.
\citet{flu99} studied the solar limb polarization between $4000$ and 8000\AA\ and found that the above expression matches reasonably well both models and observations.
Values of $\pl$ and $k$ for three wavelengths are listed in Table~\ref{tab2}.

For the later spectral types, whose limb polarization is not known, we use Eq.~(\ref{limbp}) and explore different values of $\pl$ and $k$ to study how the polarization signature varies with these parameters. 

\subsection{G Dwarfs (the Sun) \label{Sun}}

In the models shown in this and the following subsections we investigate planet radii ranging from 0.09 to 2 Jupiter radii, orbiting the star with a circular orbit of radius 0.04 AU, a typical distance for the so-called hot Jupiters \citep{sil05}. 
We point out that the only consequence on the following results, which arise from this choice of planet distance, is in the time-scale of the transit.

\begin{figure}
  \plotone{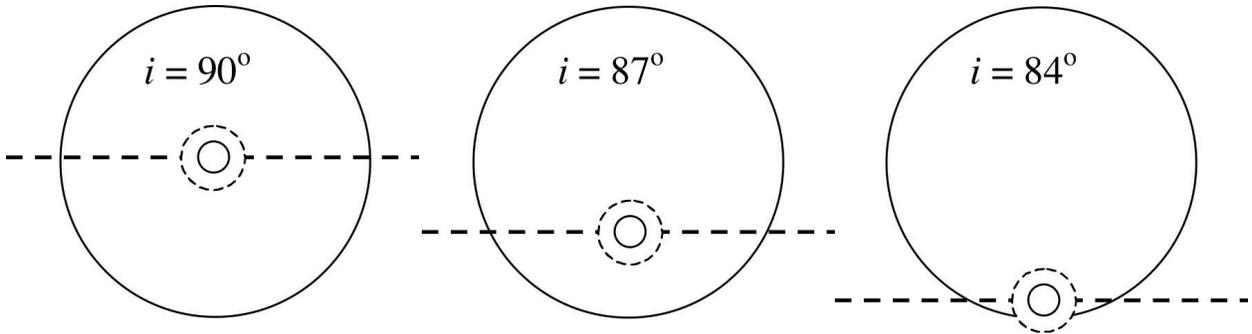}
   \caption{Geometry of the transit of a $1\rj$ ESP (solid circle) and a $2\rj$ ESP (dashed circle) in front a Sun-like star, for three different inclinations. For comparison, the geometry is equivalent to $0.5\rj$ and $1\rj$ ESPs transiting a M2 dwarf and to $0.2$ and $0.4\rj$ ESPs transiting a brown dwarf.}
    \label{transit}%
    \end{figure}

The transit of ESPs with radii $1\rj$ and $2\rj$ in front of a Sun-like star is depicted in Fig.~\ref{transit} for three different inclination angles. The figure illustrates the size of the planet compared to the star as well as the path of the transit across the stellar disk for the adopted inclinations.

\begin{figure}
  \plotone{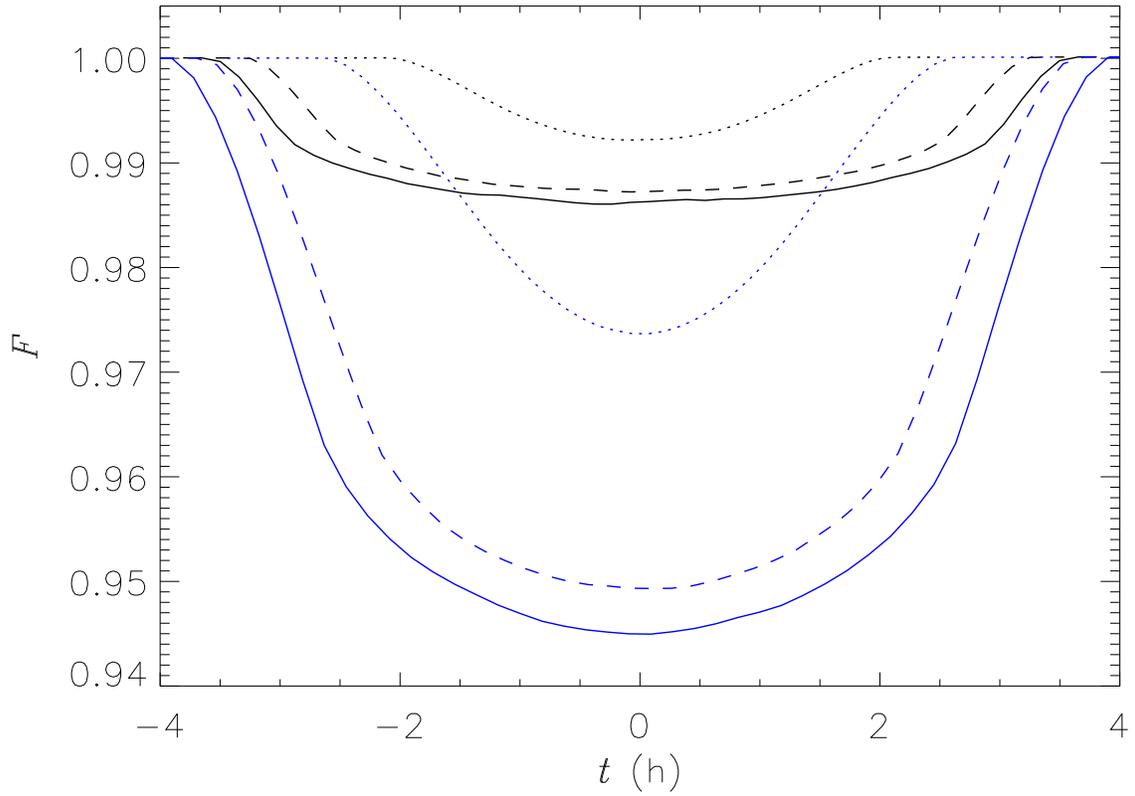}
   \caption{Normalized flux at 4600 \AA\ for the transit of an ESP in front of a Sun-like star. Shown are results for two planet radii, $R_\mathrm{p}=1\rj$ (black) and $R_\mathrm{p}=2\rj$ (blue). The solid lines correspond to central transits ($i=90^\circ$), the dashed lines to an inclination of  $i=87 ^\circ$, and the dashed lines to an inclination of  $i=84 ^\circ$. The planet orbital radius is 0.04 AU.}
    \label{sol_F_4600}%
    \end{figure}

\begin{figure}
  \plotone{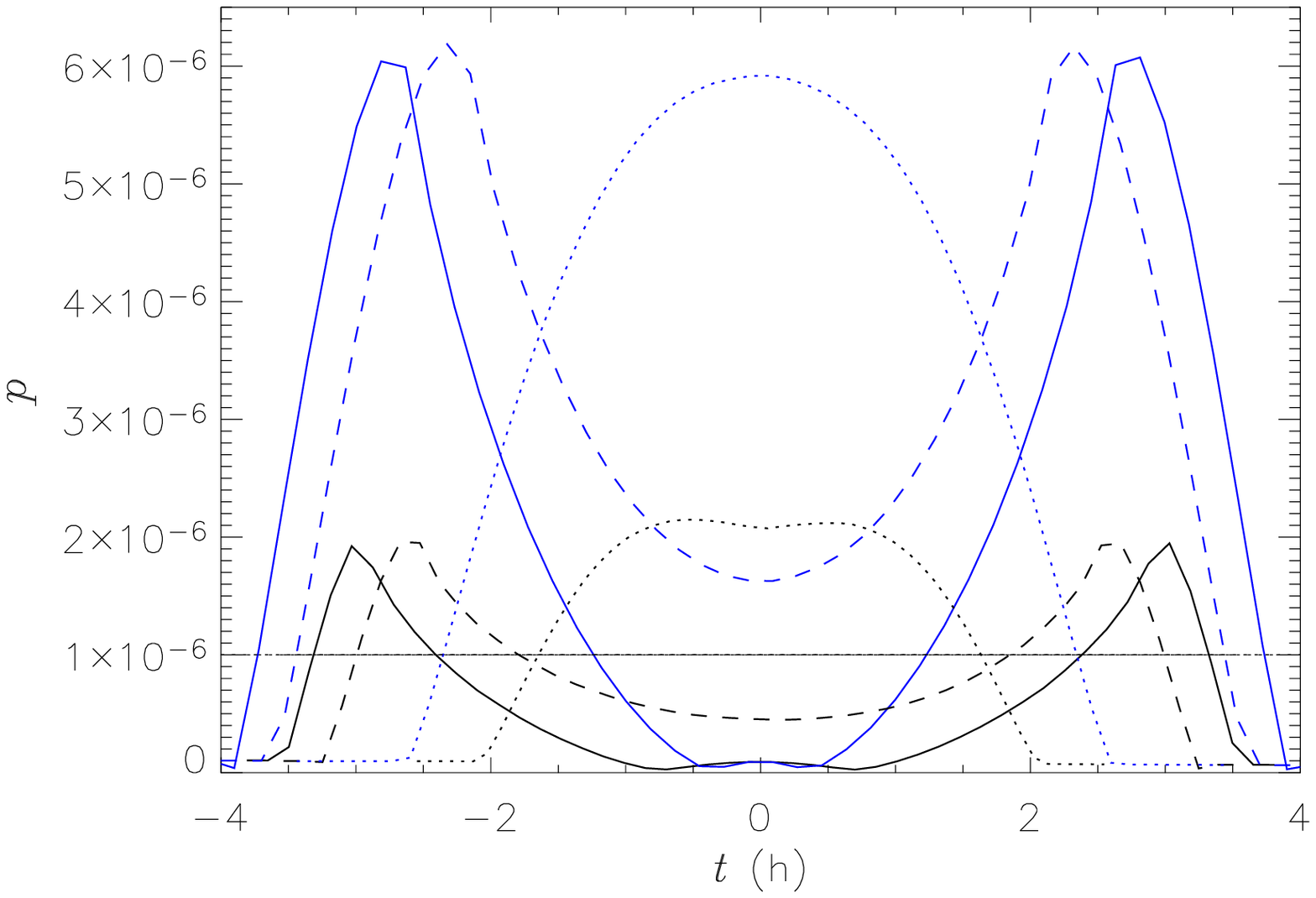}
   \caption{Occultation polarization at 4600\AA\ for the transit of an ESP in front of a Sun-like star. Shown are results for two planet radii, $R_\mathrm{p}=1\rj$ (black) and $R_\mathrm{p}=2\rj$ (blue). The solid lines correspond to central transits ($i=90^\circ$), the dashed lines to an inclination of  $i=87 ^\circ$, and the dashed lines to an inclination of  $i=84 ^\circ$. The planet orbital radius is 0.04 AU. The horizontal line indicates the best polarization sensitivity for current broad-band polarimeters.}
    \label{sol_4600}%
    \end{figure}

The monochromatic normalized flux and OP at 4600\AA\ as a function of time, resulting from the geometry of Fig.~\ref{transit}, are shown in Figs.~\ref{sol_F_4600} and~\ref{sol_4600}, respectively.
The dashed horizontal line in Fig.~\ref{sol_4600} indicates the best polarization sensitivity for current or planned broad-band polarimeters \citep[$\approx 10^{-6}$,][]{hou03}.

From Figs.~\ref{sol_F_4600} and~\ref{sol_4600} it is evident that the effects of inclination angle are quite different for the flux than for the polarization. For the flux (Fig.~\ref{sol_F_4600}), decreasing the inclination angle, which shifts the transit path on the stellar disk to higher latitudes, makes the flux curve shallower as a result of the limb darkening. Observationally, this makes photometric transit detections more difficult for smaller inclinations.

In constrast, Fig.~\ref{sol_4600} shows that the shape of the OP changes dramatically with inclination angle; it goes from a double-peaked profile for $i=90^\circ$ and intermediate inclination angles to a (nearly) single-peaked function for lower inclination angles. 
This happens because, as discussed in Sect.~\ref{ana}, the maximum OP occurs near the interior contacts of either the ingress and egress phases, for which the projected planet disk is close to the stellar limb. 
For the same reason, the maximum OP is not dependent on the inclination, provided the angle is not so small that only part of the planet disk transits in front of the star.
Halfway in time accross a central ($i=90^{\circ}$) transit, the polarization is zero because the projected stellar disk becomes symmetrical, which does not happen for other inclination angles. 

The above has two important observational consequences. First, the fact the the maximum OP is independent on the inclination means that low inclination angles will not hamper the OP detection, which may not be the case for photometric observations. Second, the strong dependence of the OP curve with inclination angle suggests polarization as a good indicator for this parameter.

As far as polarization goes, the results of Fig.~\ref{sol_4600} represent the most unfavorable scenario we explore in this paper, because of the large size of the star compared to the size of the planet and the low values of the limb polarization. Even so, the polarization levels in Fig.~\ref{sol_4600} indicate that the OP ensuing from the transit of Jupiter-like or larger planets in front of G dwarfs should be measurable. 
Larger OPs for G Dwarfs are expected in the bluer regions of the spectrum, because the values of $\pl$ increases fast with decreasing wavelength (Leroy 2000). For instance, at $\lambda=3000$\AA\ $\pl$ is about 10 times larger than at 4600\AA, and the OP in this spectral region would, thus, be proportionally larger.

\begin{figure}
  \plotone{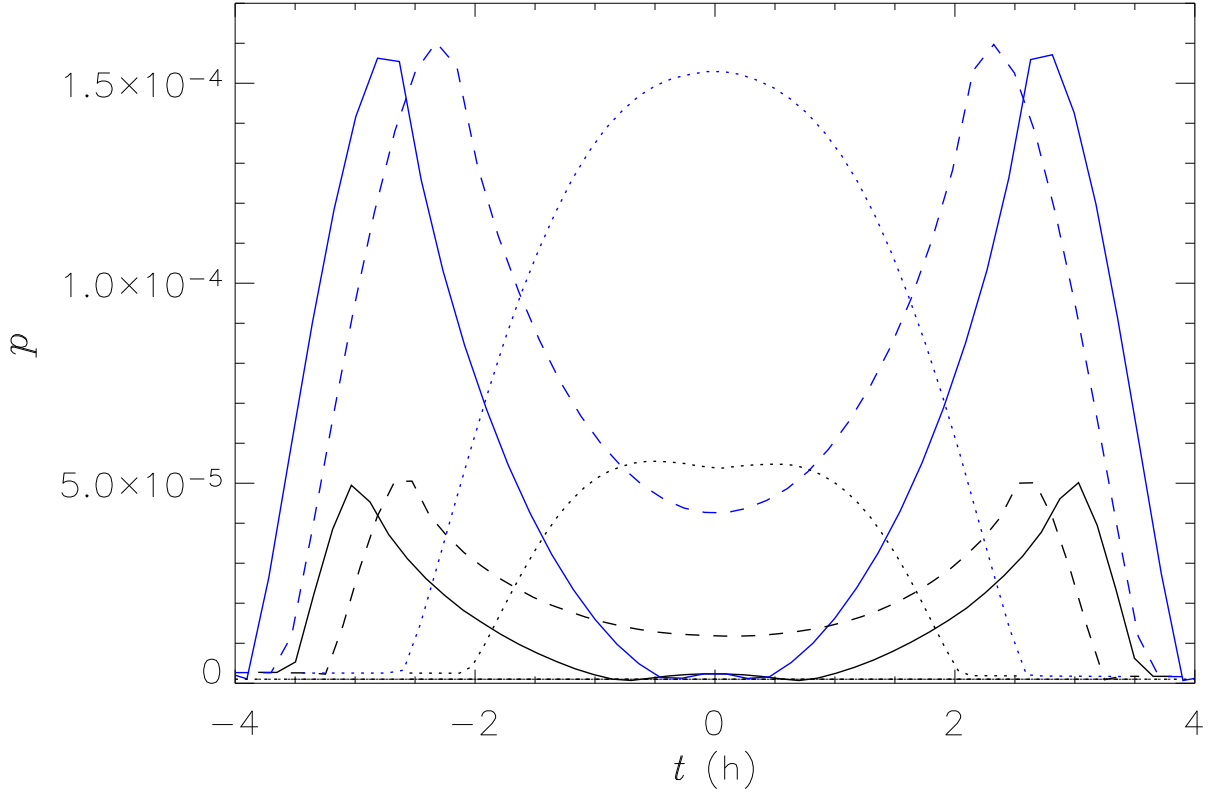}
   \caption{Same as Fig.~\ref{sol_4600} for the resonant line Ca I 4227\AA.}
              \label{sol_Ca}%
    \end{figure}

Large OPs are also expected at the wavelengths of the resonant lines of several elements, such as Ca and Sr. For instance, \citet{bia99} measured the solar limb polarization for the resonant line Ca I 4227\AA\ and found the limb polarization to be quite large ($\pl = 16.5\%$). 
Using these data, we constructed the OP for this wavelength and show it in Fig.~\ref{sol_Ca}. It can be seen that it is indeed large and well above the detection threshold.


\subsection{K, M, and T Dwarfs}

In this subsection we investigate the OP for later spectral types, for which higher polarization levels are expected. This is due to both the larger ratio between the planet and stellar radii and the higher limb polarization levels expected for those spectral types \citep{har69}. 
Since we do not know the limb polarization in these cases, we define the quantity
\begin{equation}
\epsilon \equiv \frac{p}{\pl},
\end{equation}
which can be regarded as a \emph{polarization efficiency} of the star-planet system. Given the adopted detection threshold of $10^{-6}$ and a value of $\pl$, it follows that the OP should be detectable if $\epsilon \gtrsim 10^{-6}/\pl$. 
Based on the observed solar values (Table~\ref{tab2}), a typical minimum value for $\pl$ is $\sim 0.001$, and $\epsilon$ should then be typically greater than about $0.001$ if an OP is to be measured.

\begin{figure}
  \plotone{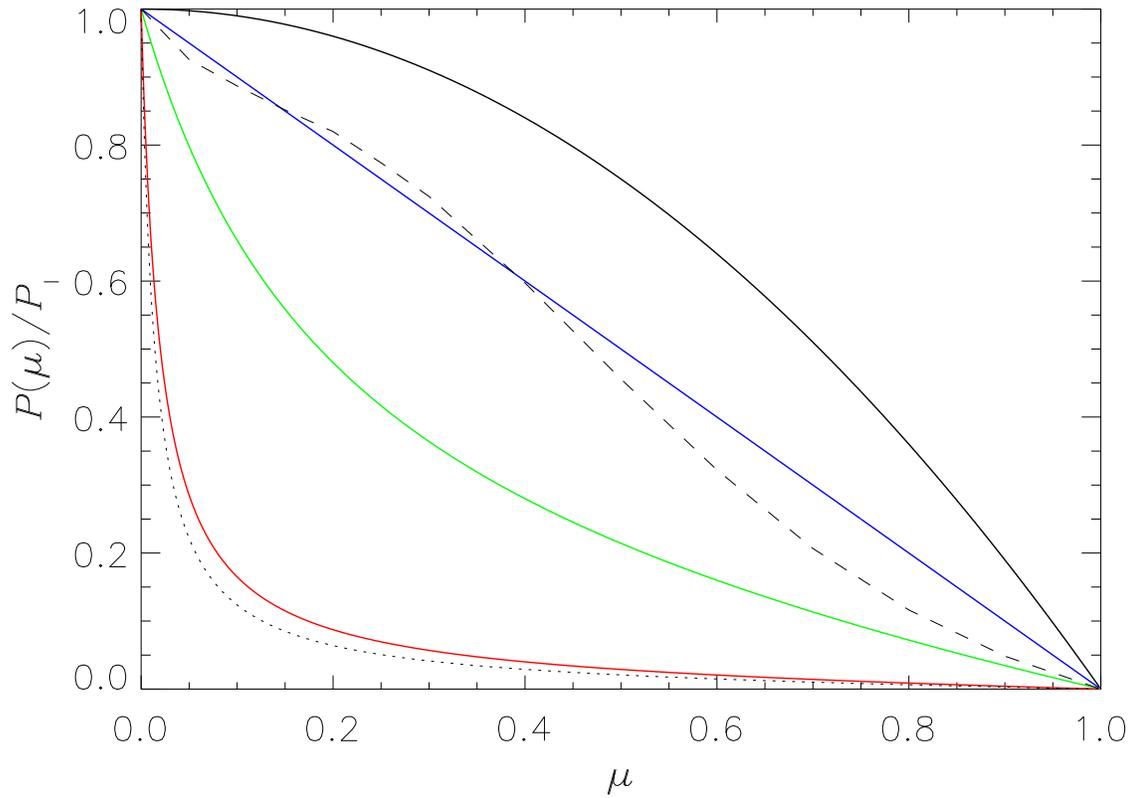}
   \caption{Limb polarization for different values of $k$. Black: $k=0$; blue: $k=1$; green: $k=5$; and red: $k=50$. The dotted line represents the curve for the solar limb polarization at 4600\AA\ and the dashed line represents the limb polarization for a M giant \citep{har69}.}
              \label{PK}%
    \end{figure}

In Fig.~\ref{PK} we first illustrate how the limb polarization curve of Eq.~(\ref{limbp}) varies with $k$. Large values of $k$, representative of the solar values (Table~\ref{tab2}), result in a very sharp decrease of the limb polarization. 
For $k=50$, for instance, the polarization drops to 10\% of $\pl$ about 0.02 stellar radii away from the limb. 
For late-type giants, the limb polarization curve is probably associated with low values of $k$ ($k\approx1$, see caption to Fig.~\ref{PK}).

\begin{figure}
  \plotone{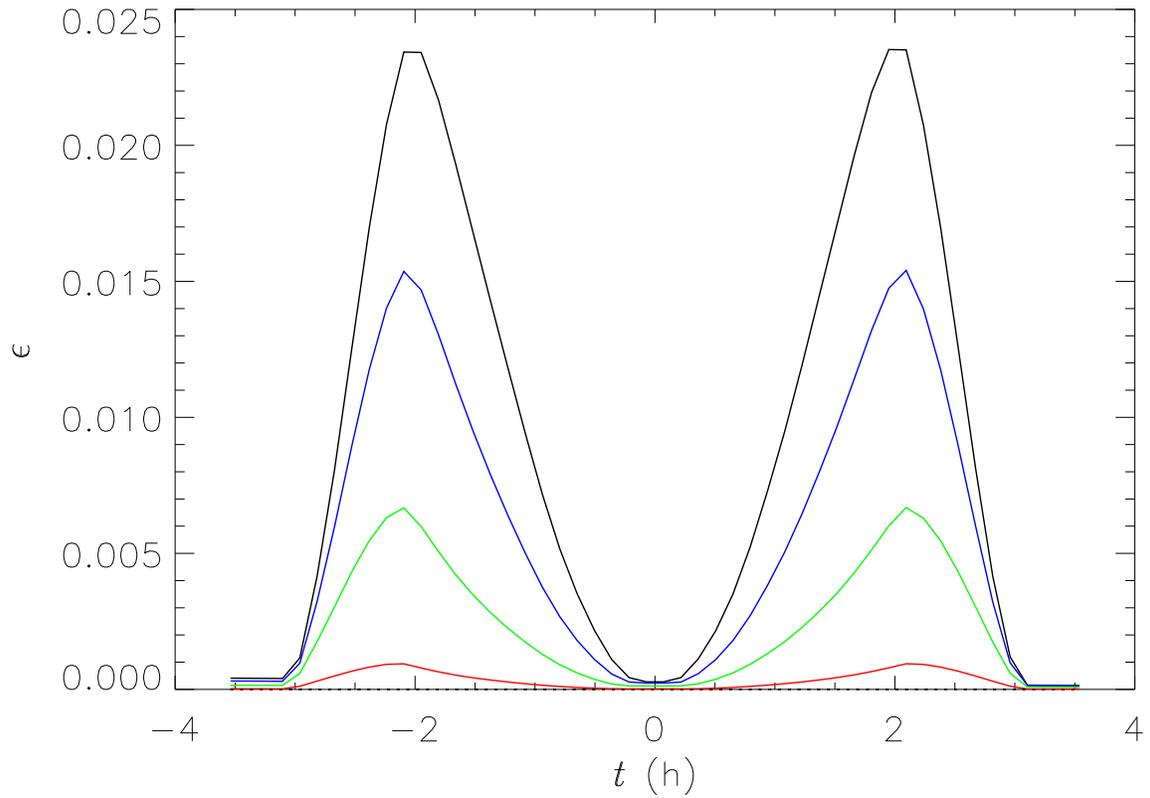}
   \caption{Occultation polarization efficiency for the central transit of a $1\rj$ planet accross a M2 dwarf. Black: $k=0$; blue: $k=1$; green: $k=5$; and red: $k=50$. For $\pl \approx 1\%$, the OP should be measurable if $\epsilon \gtrsim 0.001$.} 
              \label{K5_1}%
    \end{figure}

In Fig.~\ref{K5_1} we show the polarization efficiency $\epsilon$ for the central transit of a $1\rj$ planet in front of a M2 dwarf, for which we adopted a radius of $0.5\;R_{\sun}$ (Table~\ref{tab1}). 
The figure shows that the parameter $k$ has a dramatic effect on the polarization efficiency. 
For $k$ between 0 and 5 $\epsilon$ is quite large, resulting in observable OP levels $(\epsilon\pl)$ even for low ($\approx1\%$) $\pl$. For $k=50$ the efficiency is much smaller, but still large enough to produce OPs greater than the detection threshold, for large planets.

The explanation for the strong dependence of $\epsilon$ with $k$ is straightforward. As discussed in Sect.~\ref{ana}, the peak of the OP occurs near the inner contact of either the ingress or egress phases, and is controlled by the limb polarization at the impact parameter of the planet center at the contact. From Eqs.~\ref{pmax} and \ref{limbp} we can write the maximum efficiency as
\begin{equation}
\epsilon_{\mathrm{max}} \approx \bw^2 f(1-\bw) \frac{(1-\bw)^2}{1+k\sqrt{1-(1-\bw)^2}} \approx 
\bw^2 f(1-\bw)  \frac{1-2\bw}{1+k\sqrt{2\bw}}. \label{emax}
\end{equation}
It is evident from the above expression that the maximum efficiency drops with increasing $k$.

\begin{figure}
  \plotone{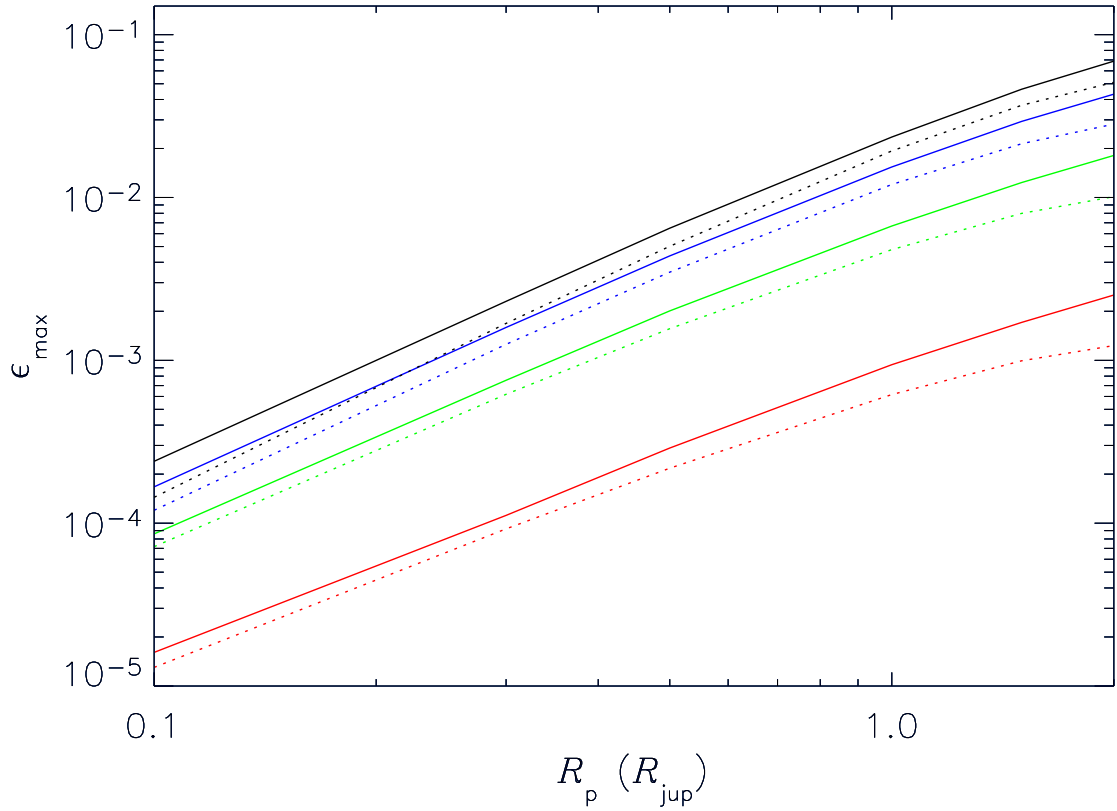}
   \caption{Maximum polarization efficiency vs. planet radius for a M2 dwarf. Black: $k=0$; blue: $k=1$; green: $k=5$; and red: $k=50$. The dotted lines represent the analytical approximation of Eq.~\ref{pmax}, with the appropriate values of $\bw$, $k$ and $a_i$.}
              \label{K5_2}%
    \end{figure}

In Fig.~\ref{K5_2} we show how the maximum polarization efficiency varies with the planet radius. 
The figure illustrates well the two opposite effects that control the OP, as discussed in Sects.~\ref{ana} and~\ref{Sun}.
Increasing the planet radius enlarges the flux absorbed by the planet and makes the polarization go up. This increase, however, is hampered by the fact that the limb polarization at the impact parameter of the planet center at the inner contact decreases with increasing planet radius. 
This is the reason why the curves in Fig.~\ref{K5_2} level off somewhat as $R_\mathrm{p}$ increases.  
In any case, all the polarization values implied by the efficiencies shown in Fig.~\ref{K5_2} should be detectable.

The dotted lines in Fig.~\ref{K5_2} represent the analytical approximation for the maximum OP, Eq.~\ref{pmax}, with the appropriate values for $\bw$, $k$ and the limb darkening coefficients.
The differences between the analytical and numerical results are small, which indicates that Eq.~\ref{pmax} is very useful for providing an estimate of the maximum OP for a given configuration.

\begin{figure}
  \plotone{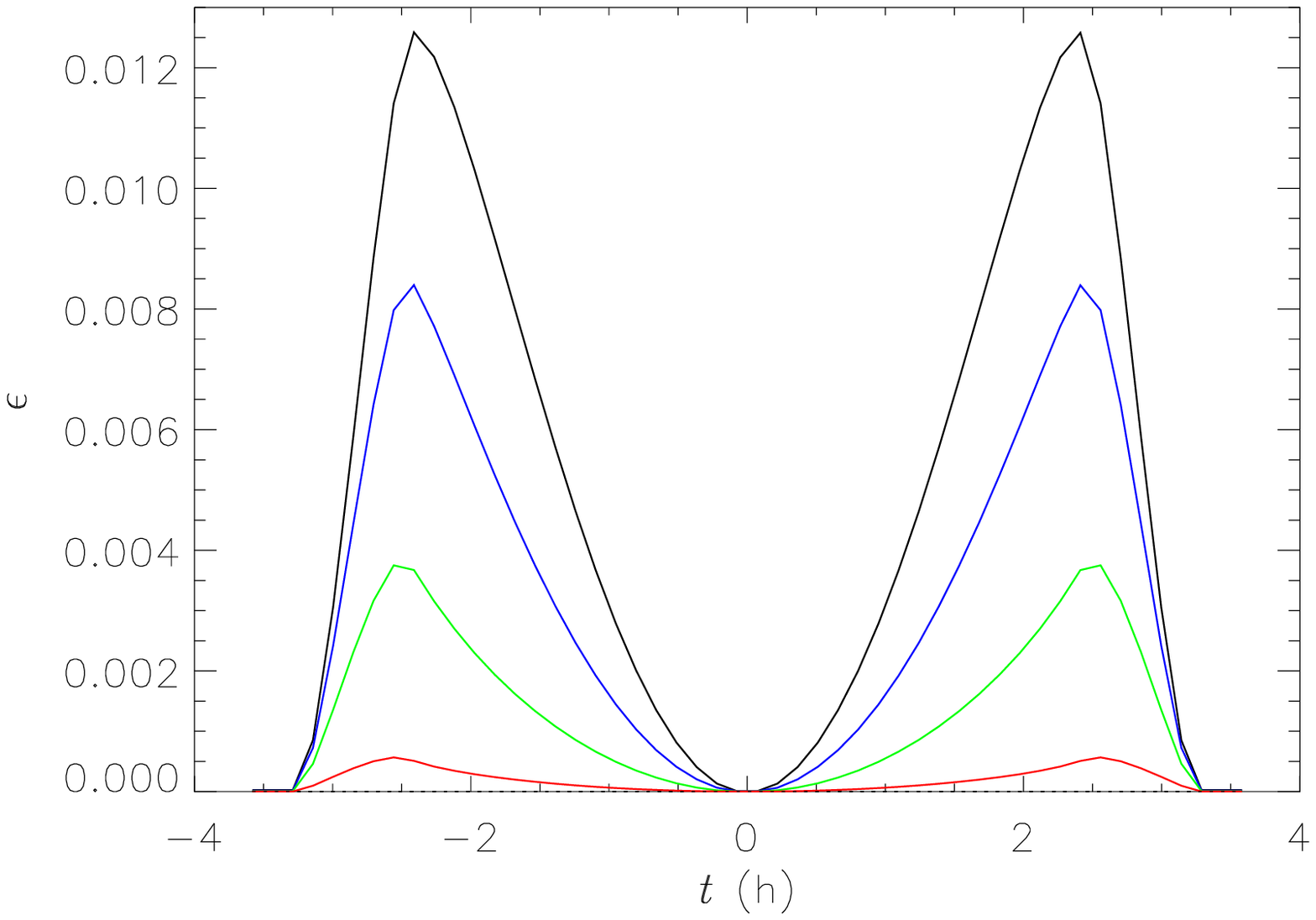}
   \caption{Same as Fig.~\ref{K5_1} for a K5 dwarf.}
              \label{M2_1}%
    \end{figure}

\begin{figure}
  \plotone{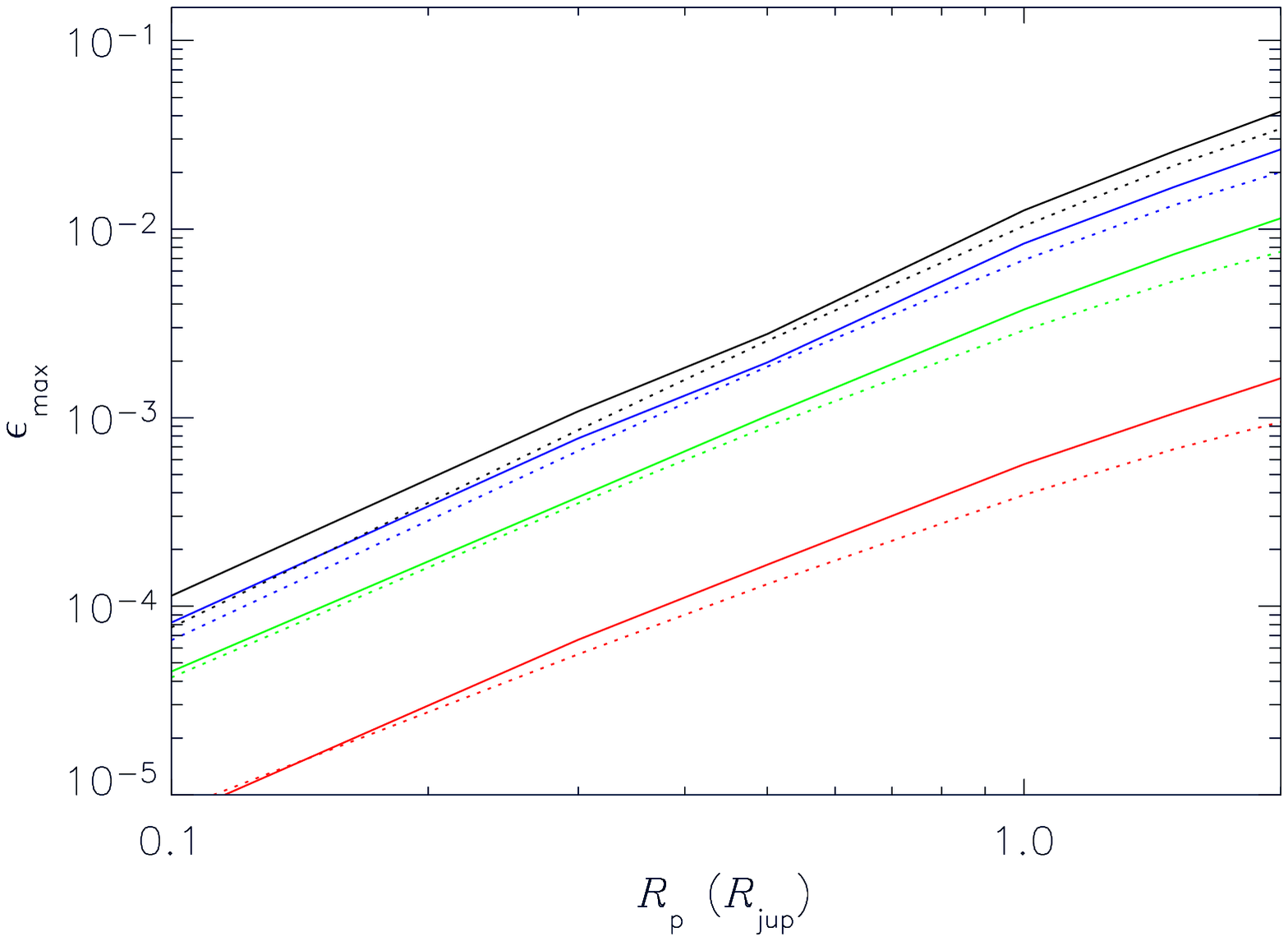}
   \caption{Same as Fig.~\ref{K5_2} for a K5 dwarf.}
              \label{M2_2}%
    \end{figure}

\begin{figure}
  \plotone{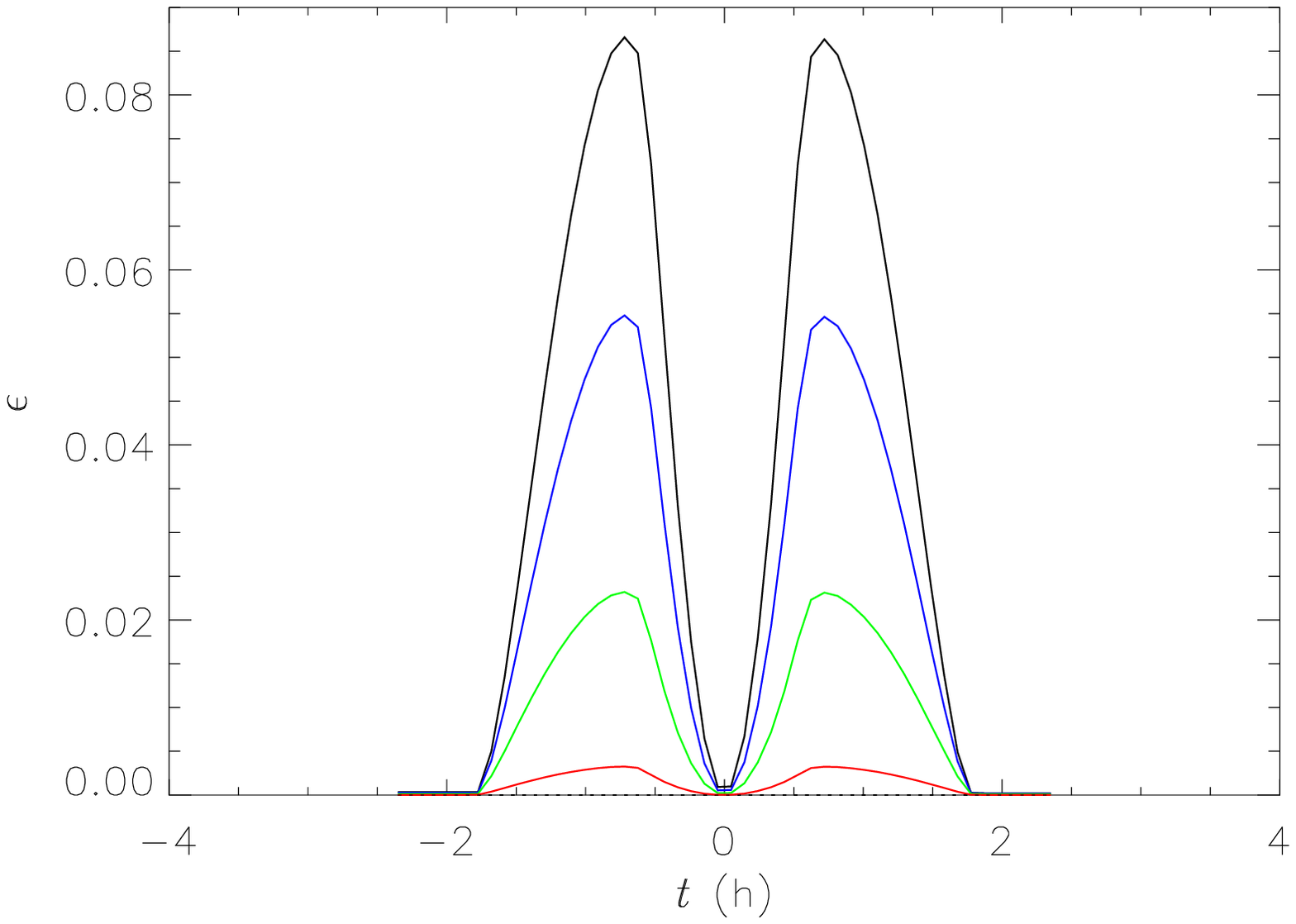}
   \caption{Same as Fig.~\ref{K5_1} for a T dwarf.}
              \label{BD_1}%
    \end{figure}

\begin{figure}
  \plotone{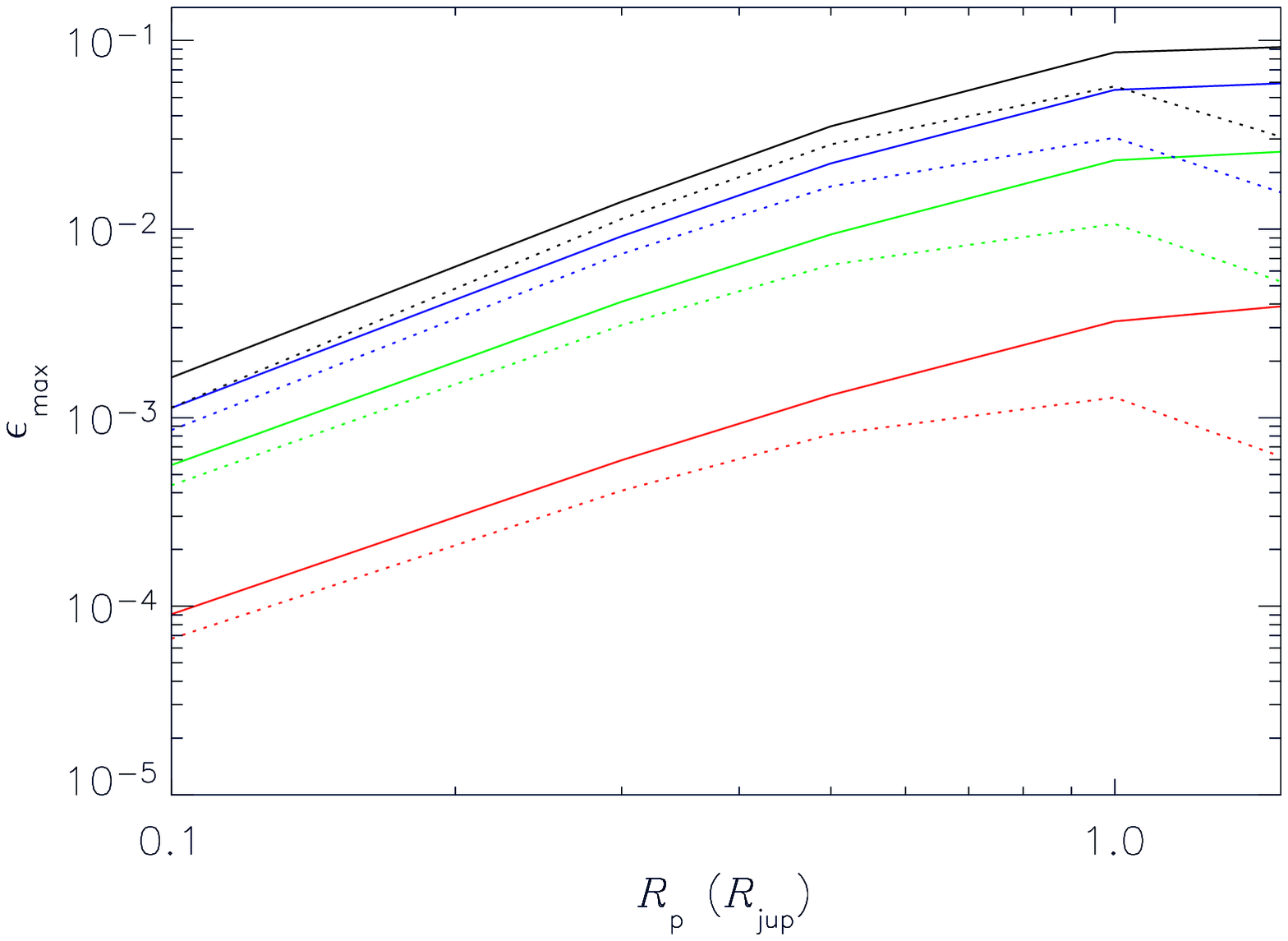}
   \caption{Same as Fig.~\ref{K5_2} for a T dwarf.}
              \label{BD_2}%
    \end{figure}

In Figs.~\ref{M2_1} and \ref{M2_2} we show the results for a K5 dwarf, for which we adopted a stellar radius of $0.72\;R_{\sun}$. The results for a T dwarf with radius $0.2\;R_{\sun}$ are shown in Figs.~\ref{BD_1} and ~\ref{BD_2}. 
We note that theoretical atmosphere models for brown dwarfs \citep[e.g.,][]{tsu02} indicate the formation of dust grains in such cool atmospheres. Those grains will provide an additional scattering opacity to the Rayleigh opacity, suggesting large values for the limb polarization for those objects. Such limb polarization needs further detailed modeling.

Our results clearly demonstrate that the OP is within the reach of current polarimeters. In the worst scenario, corresponding to the red lines of Figs.~\ref{K5_1}  to \ref{BD_2} ($k=50$), the efficiency ranges between about $10^{-5}$ ($R_\mathrm{p} = 0.09\rj$)  to 0.003 ($R_\mathrm{p} = 2\rj$),  which corresponds to polarization levels of $10^{-7}$ to $3\times10^{-5}$ if the $\pl=1\%$, and ten times those values if $\pl=10\%$.
The best scenarios correspond to $k=0$ (black lines). In these cases the efficiency is between $3\times10^{-4}$ and about 0.1. For a limb polarization of 10\%, this would correspond to polarization levels of $3\times10^{-5}$ up to 1\%.
The maximum OP values for the limiting cases studied here are summarized in Table~\ref{tab3}, where we adopted $\pl =10\%$.

Table~\ref{tab3} demonstrates that high precision polarimetry can be a means not only to study planetary systems with large (Jupiter size) planets, but also to detect Nepture-like planets ($R_\mathrm{p} \approx 0.4\rj$), and possibly Earth-like planets, depending on the shape and level of the limb polarization.

We also point out that the OP curves, exemplified by Figs.~\ref{sol_F_4600}, \ref{sol_4600}, \ref{K5_1}, \ref{M2_1} and~\ref{BD_1},  are symmetric with respect to time because of the circular orbit adopted. 
An elliptical orbit will result in an OP time dependence which will be non-symmetrical with respect to mid-transit, the details of which will depend on orbital parameters such as the orbital major-axis orientation and eccentricity. Thus, observations of the OP may potentially provide important additional orbital information which may be difficult or impossible to obtain from the light curve of the event alone.  In particular, knowledge of the eccentricity values for a sample such as the one that might result from the COROT mission should provide further insight into the dynamical evolution of the exoplanet systems thus observed.

An important point to consider is the temporal resolution that may be attainable for high precision polarimetric measurements.  
For a given absolute error $\sigma_{\rm p}$, one requires typically 
$\approx 2/\sigma_{\rm p}^2$ photons. 
For a 9th magnitude star, for instance, about an hour of exposure time on a 4 meter telescope will be needed to reach $\sigma_{\rm p}\approx 10^{-6}$. 
As a result, the adopted limit of $10^{-6}$ can only be reached for long period transits (several hours to days) or for very bright stars. 
For short transits ($\approx$ 1 hour) and/or fainter stars the detection threshold for measuring polarization variability across the event may be larger than  $10^{-6}$. On the other hand, as we pointed out above, transits that involve cool stars and/or large planets may result in OP values well above $10^{-6}$. Also, 
experiments such as the COROT satellite will likely detect long-duration transits.

We want to briefly discuss another possible source of variable polarization that might be misinterpreted as a planetary transit. Stellar spots may mimic a planet transit by removing part of the expected signal from some regions of the stellar disk, which breaks the spherical symmetry over the stellar disk, producing a polarization signal that could be similar to that of a transiting planet. 
Despite these anticipated similarities, it may be possible to separate any long-lived-spot polarization signature from the OP with detailed modeling, which is outside the scope of this paper. For instance, the net polarization as the spot moves into the observed hemisphere will likely depend on the temperature contrast between the spot and the underlying photosphere, as well as the detailed limb polarization from each location. Additionally, the spot may be intrinsically polarized by magnetic mechanisms \citep{saa03}. Finally, it might be possible to observationally distinguish the two phenomena, spots versus oculting planets, from detailed differences in their respective photometric (and possibly spectroscopic) variations.

Finally, tidal distortion induced by the planet on the star will originate a polarization signal that will vary with a period identical to the orbital period with an expected level of  $\approx 10^{-7}$--$10^{-8}$ \citep{hou03}. This interesting effect is expected to be important only for massive planets and is typically lower than the polarization signal discussed in this paper.

\section{Conclusions \label{conclusions}}

In this paper we studied the occultation polarization (OP) that ensues from the transit of an extra-solar planet in front of cool dwarfs. Following the derivation of some simple but useful analytical approximations, we presented in Sect.~\ref{results} numerical results for the OP for various planet sizes and stellar spectral types ranging from G to T. 

For the spectral type G, we used limb darkening and polarization from the Sun to calculate the monochromatic OP at 4600\AA\ and for the resonant line CaI (4227\AA). 
We found that even for such a (relatively) large star and small values of the limb polarization ($8.7\times 10^{-3}$), the OP is generally large enough ($\gtrsim 10^{-6} = 10^{-4}\%$) to be observed with modern high precision polarimeters being built \citep{hou03}. As discussed in the text, larger values are expected for the OP of G dwarfs both in the bluer parts of the spectrum and in the resonant lines of several elements, for which the limb polarization is larger than in the visual continuum.

Because of the lack of knowledge of the limb polarization for the later spectral types (K to T), we adopted the formula of Eq.~\ref{limbp} and explored several values of $k$, the width of the limb polarization curve. We found that in the least favorable cases (i.e., those of large $k$) the detection of the OP of Jupiter-like planets, down to Neptune-like planets, should be possible. In the most favorable situations (small $k$), our results suggest the interesting prospect of being able to detect Earth-like planets orbiting late type dwarfs and brown dwarfs via ground-based polarimetry.

We also studied the properties of the solution with respect to the inclination angle of the observer. We found that, while flux observations of the transit may suffer from weaker signatures at smaller inclination angles (i.e., transit path along high stellar latitutes), the same does not happen for the polarization, because the maximum OP is independent of $i$. Also, we found that the shape of the OP is much more dependent on $i$ than the shape of the flux curve, suggesting that polarimetry can help better constrain this parameter. 


\begin{acknowledgements}
      This work has been supported by S\~ao Paulo State Funding Agency FAPESP (grants 01/12589-1 and 04/07707-3). AMM acknowledges support by the Brazilian agency CNPq. The authors gratefully acknowledge professors Sylvio Ferraz-Mello and Tatiana Michtchenko for valuable discussions.
The authors thank the referee Phillip Lucas for helpful comments.
\end{acknowledgements}

  
\begin{deluxetable}{l c c c c c c}
\tablecaption{Limb Darkening Coefficients \label{tab1}}
\tablewidth{0pt}
\tablehead{
\colhead{Spectral Type} & \colhead{$a_1$} & \colhead{$a_2$} & \colhead{$a_3$} & \colhead{$a_4$} & \colhead{$R\;(R_{\sun})$} & \colhead{$T_\mathrm{eff}$ (K)}
}
\startdata
   G2 (sun @ 4227\AA) & 0.q450 &  -0.256 & 1.286 & -0.584 & 1 &5780 \\
   G2 (sun @ 4600\AA) & 0.368 &  0.265 & 0.588 & -0.356 & 1 & 5780 \\
   K5 ($V$-band)           & 0.419  & 0.704 & -0.220 & -0.035 & 0.72 &3500 \\
   M2 ($V$-band)          & 0.396  & 0.907  & -0.402 & 0.086 & 0.5 &3000 \\
   T ($V$-band)\tablenotemark{a}      & 0.396  & 0.907  & -0.402 & 0.086 & 0.2 &3000 \\
\enddata
\tablenotetext{a}{To our knowledge there is no published data about limb darkening of T dwarfs. For this reason, we used the limb darkening coefficients of the M2 dwarf, as an approximation.}
\end{deluxetable}

\clearpage
\begin{deluxetable}{c c c l}
\tablecaption{Observed Limb Polarization Coefficients for the Sun \label{tab2}  }
\tablehead{
\colhead{Wavelength (\AA)} & \colhead{$\pl$} & \colhead{$k$} & \colhead{Reference}
}
\startdata
   4227  &  $0.165$              &   50   &  \citet{bia99} \\
   4600  &  $8.7\;10^{-3}$   &    70   &  \citet{fau01} \\
   5160  &  $3.3\;10^{-3}$   &    59   &  \citet{fau03} \\
\enddata
\end{deluxetable}

\clearpage
\begin{deluxetable}{l c c c c}
\tablecaption{Maximum OP for Representative Models of Late Type Stars, using $\pl=10\%$ \label{tab3}      }
\tablehead{
\colhead{Spectral Type} &  \colhead{$R_\mathrm{p}=1\;R_\mathrm{Earth}$} & \colhead{$R_\mathrm{p}=1\;R_\mathrm{Earth}$} &
\colhead{$R_\mathrm{p}=2 \rj$} & \colhead{$R_\mathrm{p}=2 \rj$} \\
\colhead{} &  \colhead{$k=50$} & \colhead{$k=0$} &
\colhead{$k=50$} & \colhead{$k=0$}
}

\startdata
   K5  & $7.5\times 10^{-7}$ & $1.1\times 10^{-5}$ & $1.0\times 10^{-4}$ & $2.6\times 10^{-3}$ \\
   M2  & $1.7\times 10^{-6}$ & $2.4\times 10^{-5}$ & $1.7\times 10^{-4}$ & $4.6\times 10^{-3}$\\
   T     & $9.0\times 10^{-6}$ & $1.6\times 10^{-4}$ & $3.9\times 10^{-4}$  & $9.2\times 10^{-3}$\\
\enddata
\end{deluxetable}

\end{document}